\documentclass [12pt,preprint]{aastex}
\usepackage{epsfig}

\begin{document}
\title{Observation of VHE gamma-ray emission from 
the Active Galactic Nucleus 1ES1959+650 
using the MAGIC telecope}
\author{
J.~Albert\altaffilmark{a}, E.~Aliu\altaffilmark{b}, 
H.~Anderhub\altaffilmark{c}, 
P.~Antoranz\altaffilmark{d}, A.~Armada\altaffilmark{b}, 
M.~Asensio\altaffilmark{d}, C.~Baixeras\altaffilmark{e}, 
J.A.~Barrio\altaffilmark{d}, H.~Bartko\altaffilmark{f}, 
D.~Bastieri\altaffilmark{g}
% ...
}
 \altaffiltext{a} {Universit\"at W\"urzburg, Germany}
 \altaffiltext{b} {Institut de Fisica d'Altes Energies, Barcelona, Spain}
 \altaffiltext{c} {Institute for Particle Physics, ETH Z\"urich, Switzerland}
 \altaffiltext{d} {Universidad Complutense, Madrid, Spain}
 \altaffiltext{e} {Universitat Autonoma de Barcelona, Spain}
 \altaffiltext{f} {Max-Planck-Institut f\"ur Physik, M\"unchen, Germany}
 \altaffiltext{g} {Dipartimento di Fisica, Universit\`{a}  di Padova, and INFN Padova, Italy} 
 \altaffiltext{h} {Division of Experimental Physics, University of Lodz, Poland} 
 \altaffiltext{i} {Dipartimento di Fisica, Universit\`{a}  di Udine, and INFN Trieste, Italy} 
 \altaffiltext{j} {Yerevan Physics Institute, Cosmic Ray Division, Yerevan, Armenia}
 \altaffiltext{k} {Tuorla Observatory, Pikki\"o, Finland}
 \altaffiltext{l} {Institut f\"ur Physik, Humboldt-Universit\"at Berlin, Germany} 
 \altaffiltext{m} {University of California, Davis, USA}
 \altaffiltext{n} {Dipartimento di Fisica, Universit\`a  di Siena, and INFN Pisa, Italy}
 \altaffiltext{o} {Space Research Unit, Potchefstroom University, South Africa}
 \altaffiltext{p} {Fachbereich Physik, Universit\"at Dortmund, Germany}
 \altaffiltext{q} {Institute for Nuclear Research and Nuclear Energy, Sofia, Bulgaria}


\begin{abstract}
The MAGIC Cherenkov telescope has observed 
very high energy (VHE) gamma-ray 
emission from the Active 
Galactic Nucleus 1ES1959+650 
during six hours in September and October 2004. 
The observations were carried out 
alternated with  the Crab Nebula, whose data were used
as reference source for optimizing gamma/hadron separation and for 
flux comparison.  The data analysis shows 
VHE gamma-ray emission of 1ES1959+650 with $\sim$8$\sigma$ significance,
at a time of  low activity in both optical and X-ray wavelengths. 
An integral flux above $\sim 180$~GeV of about 20\% of the Crab was obtained.
The light curve, sampled over 7 days, shows no significant variations. 
The differential energy spectrum between 180~GeV and 2~TeV
can be fitted with a power law of index   
-2.72~$\pm$~0.14. The spectrum is consistent with the slightly 
steeper spectrum seen by HEGRA at higher energies,
also during periods of low X-ray activity.
\end{abstract}

\section{Introduction}

\subsection{The VHE gamma-ray source 1ES1959+650}
The Active Galactic Nucleus (AGN) 1ES1959+650 is an X-ray 
peaking BL Lacertae object  selected from the
Einstein Medium Sensitivity Survey \citep{einst}. It is hosted by an 
elliptical galaxy at a redshift of $z=0.047$.  
According to the unified model of
AGNs, BL Lacertae objects have relativistic jets 
emerging from supermassive black holes accreting at a sub-Eddington rate,
viewed under a small angle of sight \citep{padovani}.  
With decreasing luminosity, the peak frequency of the synchrotron emission from the
relativistic jets seems to move to higher frequencies. 
The class of BL Lacs in which the
synchrotron peak lies in the X-ray regime, are thus called 
HBLs (high frequency peaked BL Lacs).
The mass of the central black hole (BH) in 1ES1959+650 has been estimated
to be $\sim$1.5.10$^8$~M$_\circ$~\citep{falomo}, i.e. close to the 
BH mass of the HBL Mkn 421, the archetype of an extragalactic 
very high energy gamma-ray ($\gamma$) source \citep{punch}.

The first VHE $\gamma$ signal from 1ES1959+650 was reported in 1998 by
the Seven Telescope Array in Utah, with a 3.9~$\sigma$ significance~\citep{seven}. 
Observing the source in 2000, 2001, and early 2002, the HEGRA 
collaboration reported only a marginal signal~\citep{hegra}.
In May 2002, the X-ray flux of the source had significantly
increased. 
Both the Whipple~\citep{holder} and HEGRA~\citep{hegra1} collaborations  
subsequently confirmed also a higher VHE $\gamma$  flux.
Further high $\gamma$ activity periods were detected in the same year,
with some flares exceeding the Crab flux by a factor 2-3. 
An interesting aspect of the source activity in 2002 was the observation 
of a so-called {\it orphan flare} 
(viz. a flare of VHE $\gamma$s not accompanied 
by correlated increased activity at other wavelengths), 
recorded on 4 June by the Whipple collaboration~\citep{multiwl,sedwhipple}. 
The HEGRA collaboration had observed another, less significant, 
orphan VHE signal during moonlight 
two days earlier~\citep{tonello1, tonello2}. Both flares in
VHE $\gamma$s, observed in the absence of high activity in X-rays,
are not expected from the synchrotron self-Compton (SSC) mechanism 
in relativistic jets \citep{kellermann}.
For other HBLs, models based on the SSC mechanism \citep{ghisellini} can
successfully explain most of the VHE $\gamma$ production. Future observations of
1ES1959+650, therefore, are of special importance.

This paper is structured as follows: after a 
brief description of the MAGIC telescope, we present
in Sect.~\ref{sect:analysis} the data analysis using image parameters for 
gamma-hadron separation, and the reconstruction of the
direction and energy of the measured photons.  Results are shown in 
Sect.~\ref{sect:results} comparing with
data from the Crab Nebula taken around the same time and
under similar zenith angles.  
Finally, we discuss in Sect.~\ref{sect:conclusion} some implications
of our findings for VHE emission models and the extragalactic background light.

\subsection{The MAGIC Cherenkov telescope}

The MAGIC telescope represents a new generation of 
Imaging Air Cherenkov Telescopes (IACTs) 
for $\gamma$ astronomy. 
Its design  has been optimized to achieve a trigger 
threshold lower than was possible with
previous IACTs (MAGIC eventually is to reach a trigger
threshold of 30~GeV at zenith). 
The low threshold will 
make it an ideal instrument for the study of VHE $\gamma$ sources 
that have spectral cut-offs below 100-200~GeV, such as 
pulsars, medium redshift AGNs, etc.

The MAGIC parameters and performance 
have been described  elsewhere~\citep{magictech, magiccomm}. 
The MAGIC mirror has a diameter 
and focal length both of 17m;
its camera comprises 576 hemispherical photo-multiplier tubes with 
diffuse lacquer coating~\citep{laquer} and specially shaped 
light collectors, both enhancing quantum efficiency.
The camera has a field of view (FOV) of 3.5$^{\circ}$.

The MAGIC telescope is located on the Canary Island of La Palma 
(28.2$^{\circ}$N, 17.8$^{\circ}$W, at 2225 m asl). 
From this location, 1ES1959+650 is visible from May to October under a zenith 
angle of 36$^{\circ}$ at culmination.
At a mean observation angle of 40$^{\circ}$, 
the threshold for the physics analysis is 
about twice that at zenith. We present here an analysis down to 180~GeV. 
Past Whipple and HEGRA observations were carried out above 700~GeV 
and above 1~TeV, respectively.

\section{\label{sect:analysis}Data Analysis}

The analysis presented here is restricted to $\gamma$s 
with an energy above 180~GeV.
At such energies, we can 
discriminate hadronic and electromagnetic showers
using the classical techniques  
pioneered by the Whipple collaboration, 
described in ~\citep{fegan}.
The shower image in the camera is parameterized to obtain 
several test statistics~\citep{hillas} describing the image shape and 
orientation (also called image parameters or 
discriminant quantities).
The parameters are used to reject hadronic background events
by defining, in the space of these parameters, 
limiting values, {\it cuts},
that discriminate between $\gamma$- and hadron-induced images. 
The parameters also
permit reconstructing the arrival direction and the 
energy of the original $\gamma$.

Table \ref{tab:rawdata} shows the summary of the data collected from 
1ES1959+650, Crab Nebula, and OFF-source. This period in 
fall 2004 corresponds to the
end of the MAGIC commissioning phase.
The zenith angles for these observations are all in 
the range 36~-~46$^\circ$.

\begin{table}[htdp]
\caption{\small \it Statistics of the raw data analyzed.}
\begin{center}
\begin{small}
\begin{tabular}{|c|c|c|c|}
\hline
\textbf{Source} & \textbf{Date in 2004}  & \textbf{Total time} & \textbf{N. events}\\
\hline
\hline
1ES1959+650 & Sept. 6-7, Oct. 7, 10, 14-17  & 6 h 31 min   & 4.4 M  \\
\hline
Crab       & Sept. 13-16, 21-23            & 2 h 17 min    & 1.7 M  \\ 
\hline
OFF-source & Sept. 8, 10-13, 17           & 2 h 49 min   & 2.3 M  \\
\hline
\end{tabular}
\end{small}
\end{center}
\label{tab:rawdata}
\end{table}

Generally, the Crab Nebula with its very stable flux is considered a reference
source, viz. a standard candle, for VHE $\gamma$ astronomy. 
For that reason, Crab data observed with MAGIC were selected such as to 
match  telescope operation conditions, in time and zenith angle, 
to those during the observation of 1ES1959+650. 
So-called OFF-source data are collected by pointing the telescope to a sky 
section near the source, where no $\gamma$ 
signal is expected in the field of view. 
These data are used as crosscheck of the recorded cosmic ray background. 

After quality cuts (rejection of accidental triggers due to noise etc.), 
and correcting for the dead-time of the data acquisition system, the effective
observation time for 1ES1959+650 amounts to $\sim$6~hours. 
The optimal cut values of image parameters for the $\gamma$-hadron separation 
was obtained using Monte Carlo data\footnote{\small The MAGIC Monte Carlo programs
are based on Corsika 6.019, see \cite{corsika}.}, the parameter cut 
values being 'trained' to obtain a signal with the maximum significance 
from the $\sim$2~hours of Crab Nebula data observed at the same zenith angle.
These cuts were then applied to the 1ES1959+650 data sample, 
without further optimization. In our analysis, we used eight image
parameters\footnote{\small The parameters are ALPHA, SIZE, DIST, transformed
WIDTH and LENGTH, two different concentration parameters, 
and an asymmetry parameter.};
the optimization procedure used the {\it Random Forest} method,
which optimizes the transformation of the parameter space into a single
variable, called {\it hadronness} \citep{breiman,bock}. More details 
on the analysis can be found in~\cite{tonello2}.

Two of the image parameters are of particular importance: 
the variable  SIZE, expressed in number of photo-electrons in the camera, 
is, for an impact parameter between $\sim$50m and $\sim$150m 
(equivalent to the image parameter DIST between 0.3$^{\circ}$ and 1$^{\circ}$), 
to first order proportional to the energy of the incoming $\gamma$; 
the variable ALPHA, the angle in the image between the major axis 
and the direction of the source, shows most clearly
the existence of a signal. ALPHA is not included in 
the optimization process; instead, 
after optimizing cuts in the 
other parameters, we derive from the ALPHA distribution 
(Fig.\ref{fig:alphaplots}) the 
significance of the signal (using~\citep{lima}, formula 17).
Finally, the data are required to satisfy low ALPHA values,
thus selecting only showers that point to the source position.

\section{\label{sect:results}Results}
\subsection{Alpha plot and comparison with Crab}

\begin{figure} [htb]
\plottwo{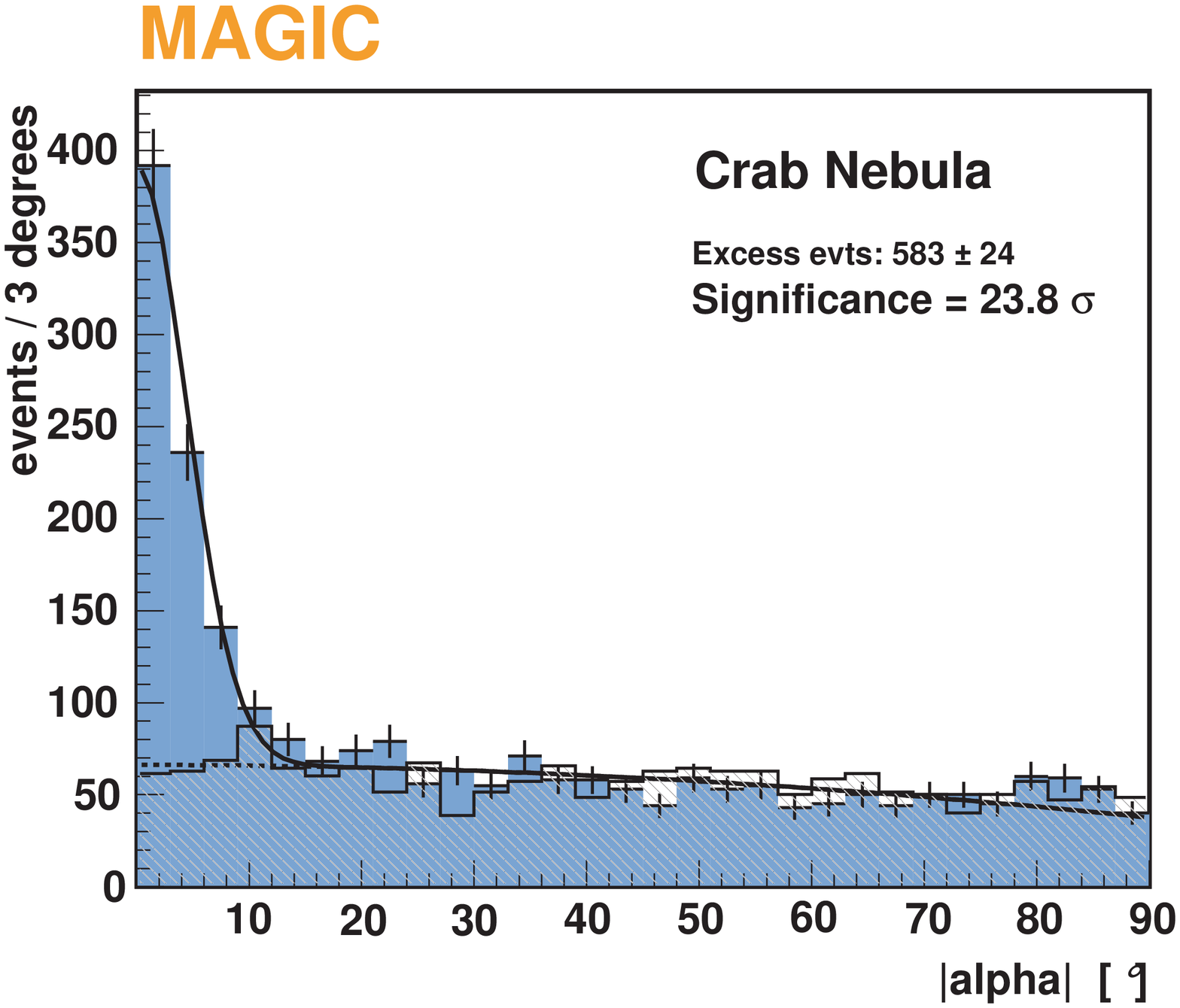}{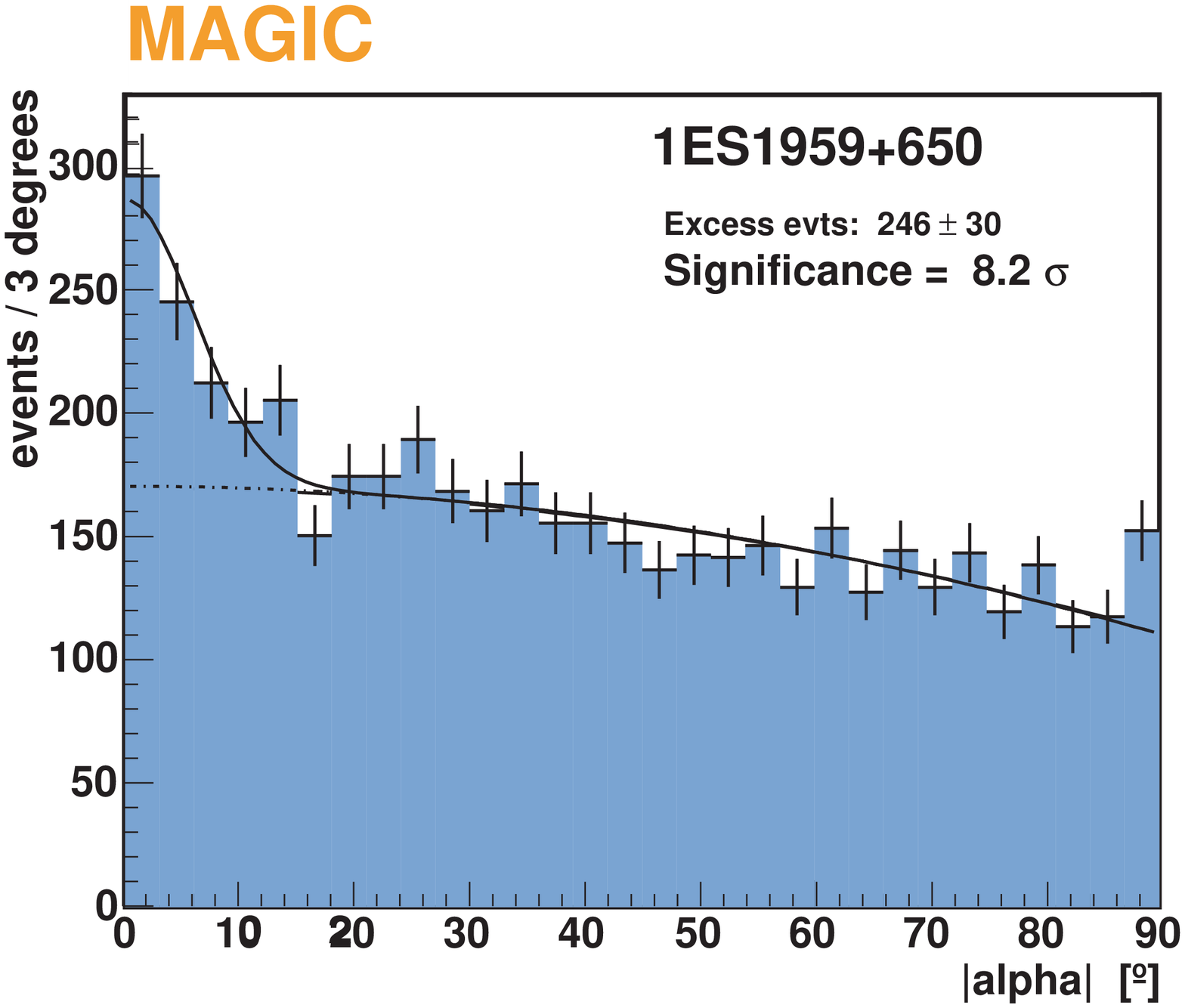} 
\caption{\small \it ALPHA plots of Crab (left)
and 1ES1959+650 (right), after cuts on image parameters. 
Both diagrams show the second-order curve 
used for estimating the background at low ALPHA (up to 9$^\circ$). 
In the left diagram (Crab), 
we also have added the (normalized) OFF-source data. 
\label{fig:alphaplots}}
\end{figure}

In Fig.\ref{fig:alphaplots} (left) we show the distribution 
of the image parameter ALPHA for the Crab Nebula, together with the 
OFF-source data normalized to the ON-source data between 
20$^\circ$ and 90$^\circ$. 
Here we chose a selection of events in terms of SIZE,
corresponding to a threshold~$>$300~GeV
\footnote{\small This data selection has been chosen such as to 
allow applying a constant ALPHA cut of 9$^\circ$
to the entire sample. Including lower energy events adds more
background and reduces significance. The optimization of the 
full sample, including events of smaller SIZE, has been done 
separately.}. 
In Fig.\ref{fig:alphaplots} (right), the ALPHA distribution 
of the 1ES1959+650 data 
sample is shown, after applying the parameter cuts optimized using the 
Crab sample.  The background for the Crab data under 
the signal was estimated
both from the OFF-source events and by 
extrapolating the ALPHA distribution 
from ON-source events between 20 and 90$^\circ$, 
using a simple second-order formula 
($C_1 + C_2\cdot ALPHA^2$). 
Both methods give the same result; we thus used the same formula
of extrapolation from the ON-source events outside the excess peak,
for both Crab and 1ES1959+650.

The significance of the 1ES1959+650 detection is 8.2$\sigma$, 
with 246$\pm$30 excess events (after all cuts)
in $\sim$6.0 hours, the signal from Crab Nebula corresponds 
to $\sim$23.8 sigma and 583$\pm$24 excess events in $\sim$2.1 hours. 

We obtain a integral VHE $\gamma$ flux 
from 1ES1959+650 above 180~GeV of
(3.73~$\pm$0.41~$\pm$0.35)~$\cdot$10$^{-11}$~photons~cm$^{-2}$~s$^{-1}$
(the errors given are statistical and an estimate for systematics, 
respectively)\footnote{\small 
For this estimate of the flux, the small correction for the 
'dead time' of the electronics readout was not considered.}. 
For the flux above 300~GeV, the result is
(1.57~$\pm$0.17~$\pm$0.30)~$\cdot$10$^{-11}$~photons~cm$^{-2}$~s$^{-1}$.
These flux values correspond to 0.20 and 0.17 Crab units, respectively, when
comparing to the Crab flux measured by MAGIC.

We also analyzed the data set using a completely independent analysis 
chain\footnote{\small using dynamical supercuts as described in \cite{kranich}},
and obtained, within statistical limits, the same significance and flux.

\subsection{The light curve}

\begin{figure}[htb]
\plotone{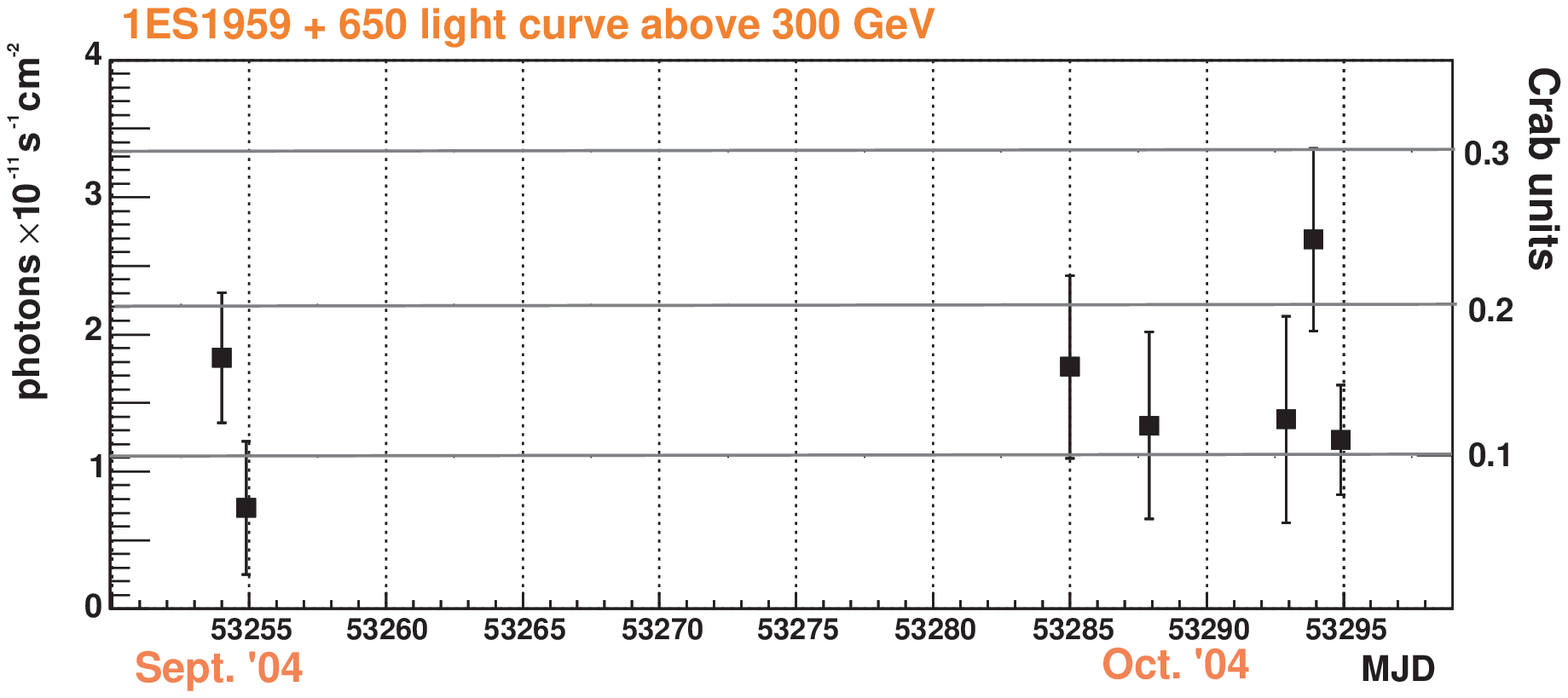}
\caption{\small \it Light curve for 1ES1959+650 as observed by 
MAGIC during seven nights
\label{fig:VHElightcurve}}
\end{figure}

Most blazars known to emit VHE $\gamma$s were detected at times of strong
VHE $\gamma$ flaring, and correlated with
strong X-ray variability during the same period.
For our 6~hours observation time of the 1ES1959+650, only modest tests 
of the flux variation are possible. 
We show in Fig.\ref{fig:VHElightcurve} and in table~\ref{tab:lightcurve}
a flux analysis for each night, indicating that 
the source was basically in the same (low) state during the time covered 
by our observation; corrections for small differences in the zenith 
angle came out to be negligible.

\begin{table}[htd!]
\caption{\small \it Analysis of 1ES1959+650 data divided into single nights 
of observation.}
\begin{center}
\begin{tabular}{|c|c|c|c|c|}
\hline
{\bf Date}& {\bf Excess}&{\bf Sign.}&{\bf Flux ($>$ 300 GeV)}\\
 MJD &   {\bf  evts/min}&  $\sigma$   &  $10^{-11}$ ph.el. cm$^{-2}$ s$^{-1}$\\
\hline
\hline
53254.0 &   0.82$\pm$0.22 &    3.7 & 1.83 $\pm$ 0.47 \\
53254.9 &   0.54$\pm$0.24 &    2.2 & 0.74 $\pm$ 0.48\\
53285.0 &   0.95$\pm$0.28 &    3.4 & 1.76 $\pm$ 0.67\\
53287.9 &   0.95$\pm$ 0.30     &3.2& 1.34 $\pm$ 0.68\\
53292.9 &   0.53$\pm$  0.31    &1.7& 1.38 $\pm$ 0.75\\
53293.9 &   1.26$\pm$  0.27    &4.7& 2.69 $\pm$ 0.67\\
53294.9 &   0.69$\pm$  0.18    &3.9& 1.23 $\pm$ 0.40\\
\hline
\end{tabular}
\end{center}
\label{tab:lightcurve}
\end{table}

\subsection{Comparison with simultaneous observations at other wavelengths}

\begin{figure}[htb]
\plotone{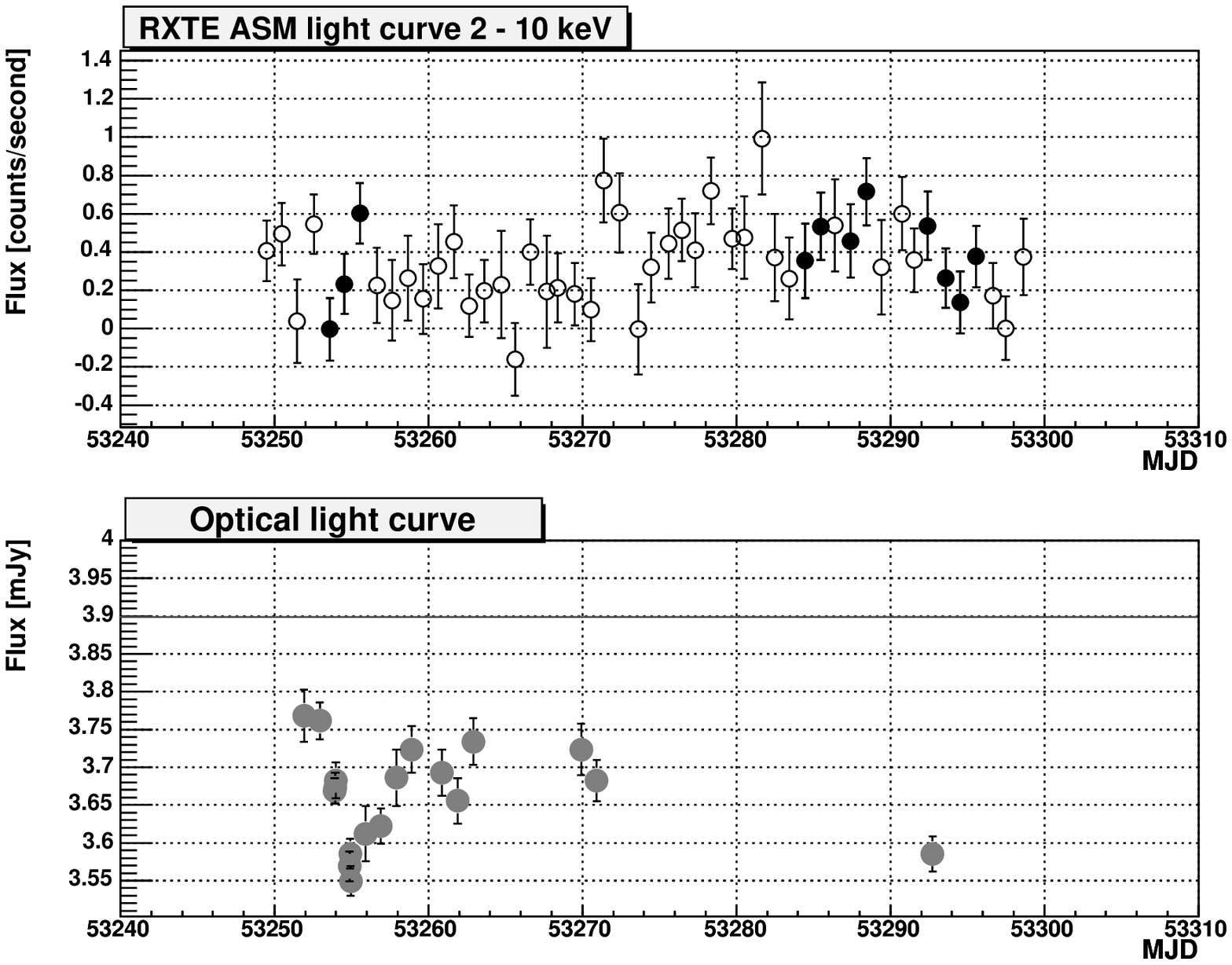} 
\caption{\small \it Top: Light curve in X-rays for 1ES1959+650 
during the months of September 
and October 2004 (from published RXTE-ASM data). 
The full circles indicate those recorded during the period of $\gamma$ 
observations with MAGIC. Bottom: Optical light curve  for the same period
(from the Tuorla Blazar Monitoring Program). The line at 3.9~mJy gives the average
flux over nearly two years (2002-09-10 to 2004-08-25) before the MAGIC observations:
during our observations, the optical activity was particularly low. 
\label{fig:Xlightcurve}}
\end{figure}

\begin{figure}[htb]
\plotone{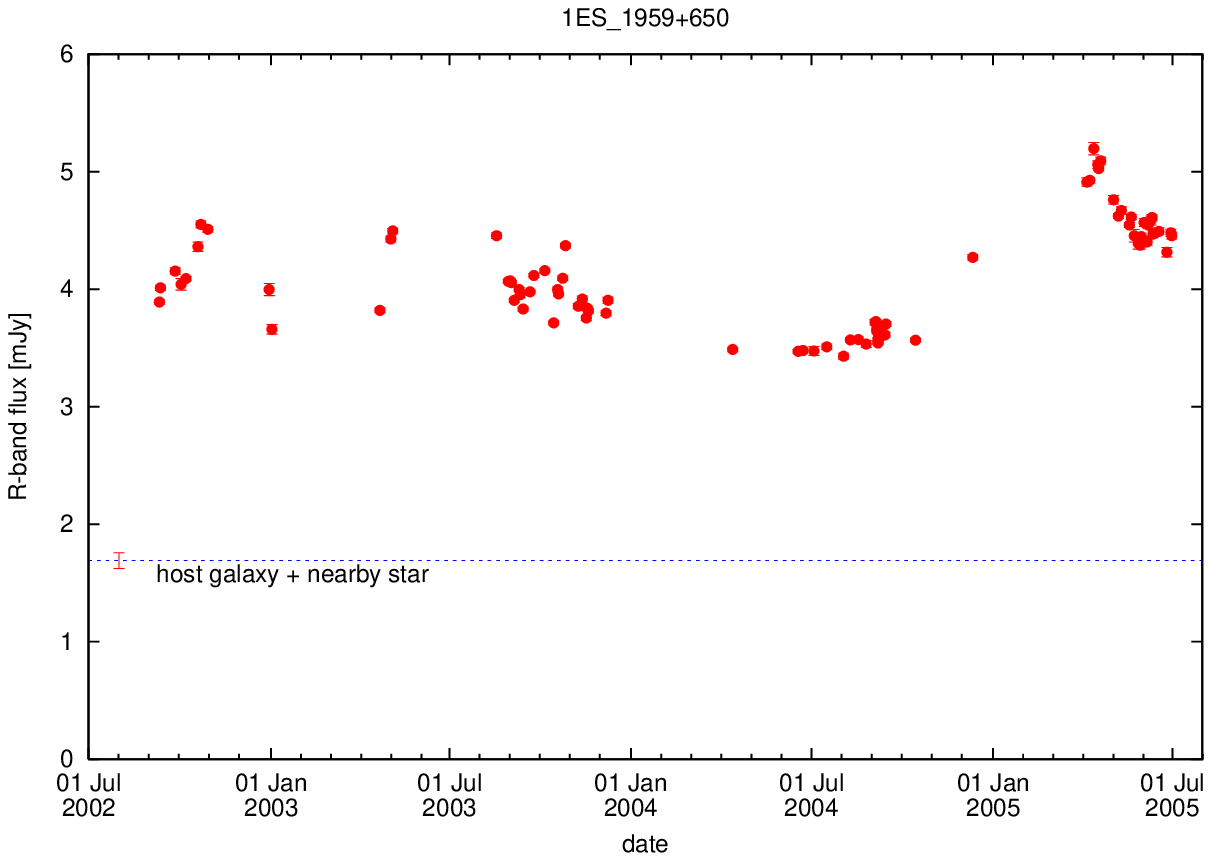} 
\caption{\small \it Full circles: light curve in optical for 1ES1959+650 
from July 2002 to January 2005 (from the Tuorla Blazar Monitoring Program). 
The luminosity of the host galaxy is also reported.
\label{fig:Olightcurve}}
\end{figure}

Strong VHE $\gamma$ emission from an AGN leads to the question whether the 
source was active also at other wavelengths.
If the $\gamma$ emission is due to the inverse-Compton
scattering of accelerated electrons, their corresponding synchrotron 
emission must show up at lower
energies.  Most observations of other sources are
indeed in line with the correlated 
X-ray variability expected from
synchrotron-self-Compton models, e.g~\citep{ssc1,ssc2}.
Figures \ref{fig:Xlightcurve} and \ref{fig:Olightcurve} show the light curves 
of 1ES1959+650 in the X-ray and optical domains. 
The X-ray data are based on published RXTE-ASM X-ray 
flux data \citep{asmdata}, the optical 
light curve is provided by the Tuorla Observatory Blazar Monitoring Program 
\citep{optdata}.
No strong activity in X-rays or in the optical was observed during the period of 
the VHE $\gamma$ studies reported here. This fact and the absence of significant time 
variability lead to the tentative conclusion 
that the reported VHE $\gamma$ emission of 1ES1959+650 does not follow the pattern 
observed in other AGNs during flaring periods. Future observations over longer 
periods will shed more light on the nature of the quiescent 
VHE emission of 1ES1959+650 (see Sect.~\ref{sect:conclusion} below).

\subsection{The VHE $\gamma$ spectrum and a comparison with the 
Crab spectrum}

\begin{figure}[htb]
\plotone{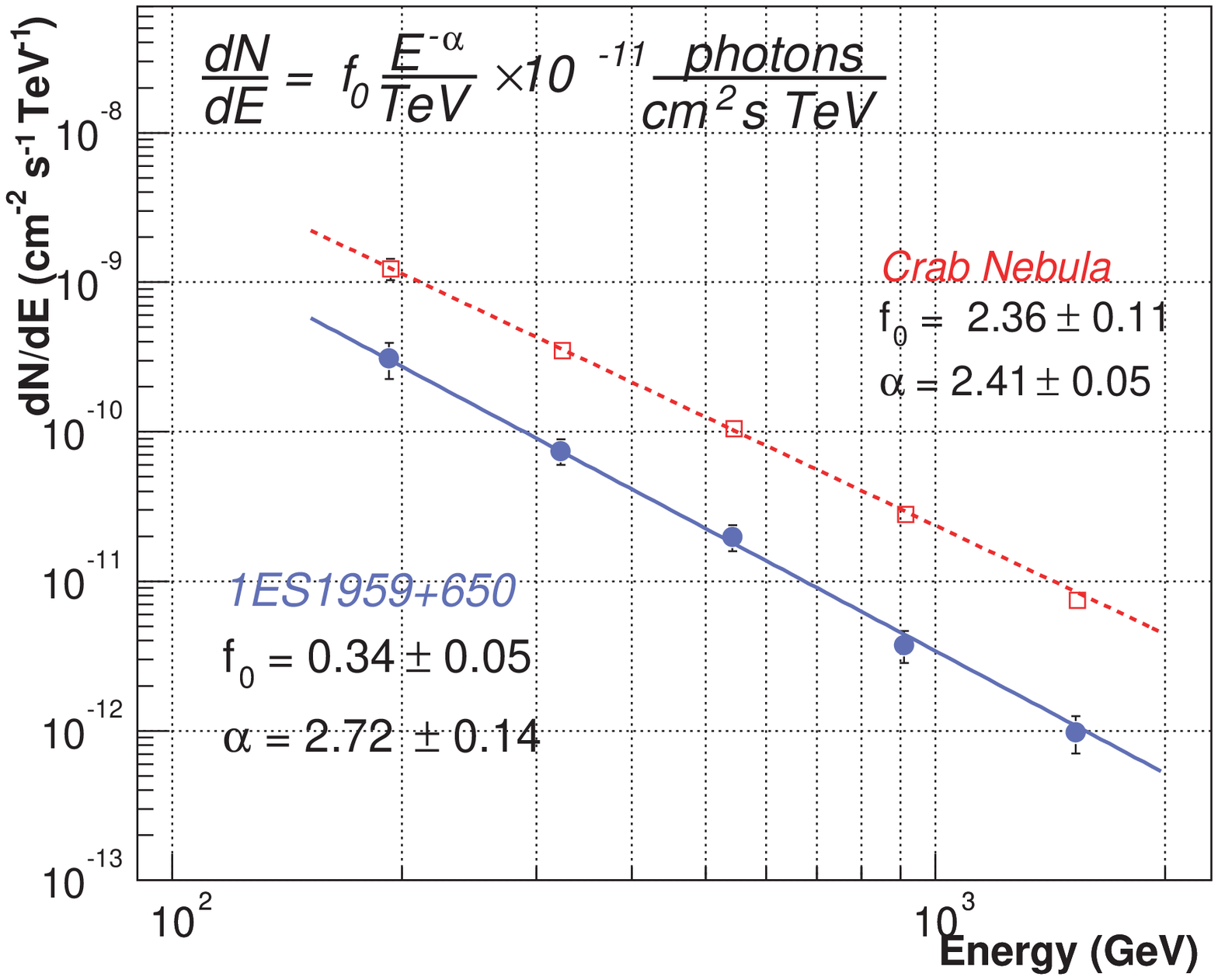}  
\caption{\small \it Differential spectra for Crab 
and 1ES1959+650. The energy range from 150~GeV
to 2~TeV is divided into five bins in logarithmic scale. The point positions
are the median values of the estimated energy bins, weighted with the
assumed spectral slope.
\label{fig:spectrum_1}}
\end{figure}

\begin{figure}[htb]
\plotone{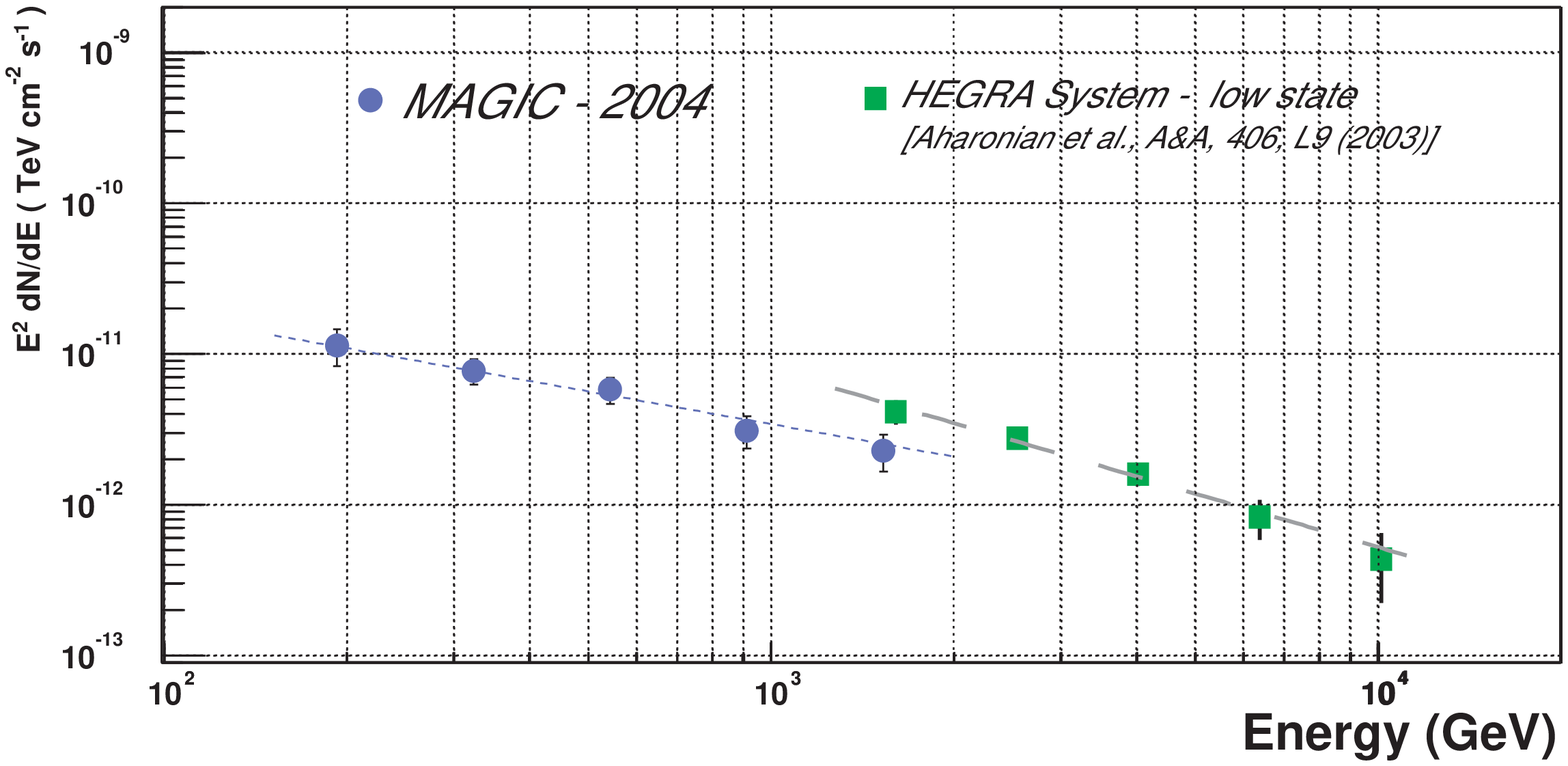} 
\caption{\small \it Differential spectrum 
for 1ES1959+650, combined with HEGRA points from \citep{hegra1}. 
Note that the HEGRA points are also from a low state of activity, but
that there is no unique definition of 'low state'.
\label{fig:spectrum_2}}
\end{figure}

The spectra for Crab and 1ES1959+650 measured in fall 2004
are shown in Fig.\ref{fig:spectrum_1}.
Both spectra are consistent with a simple power law,
albeit with a spectral index smaller than reported at higher energies.
The lines show fits to the spectra between 180~GeV and 2~TeV corrected
for spill-over factors,
using as ansatz an  unbroken power law; the fits give 
slopes of -2.72$\pm$0.14 for 1ES1959+650 and -2.41$\pm$0.05 for the Crab,
respectively.
There is strong evidence of the spectrum of the 1ES1959+650 
being steeper than that of the Crab over this energy range. 

In Fig.\ref{fig:spectrum_2} we show a comparison with spectral data taken
by HEGRA in 2002 at higher energies \citep{hegra1}, 
with a slope of -3.18$\pm$0.17.
The energy overlap of past and current data is small, demonstrating 
the progress in accessing lower energies with the MAGIC telescope.
Past spectral descriptions required a cut-off parameter of 
about 3~TeV, in order to take into account possible absorption 
due to the cosmic infrared background. 
Our data are in an energy range where the effects of such an absorption 
process are weak, thus a simple power law should be sufficient to 
describe the data.
 
\section{\label{sect:conclusion}Discussion and Conclusions}

The HBL 1ES1959+650 has been clearly detected with the MAGIC 
telescope, in a few hours observation time during September 
and October 2004, at a mean zenith angle of~40$^{\circ}$.
During that period, the source was in a quiescent state both 
in X-rays and at optical wavelengths. 
In the same period, Crab Nebula and OFF-source data were recorded 
under comparable observational conditions.  

For the first time, 1ES1959+650 has been observed down to 180~GeV, 
a limit much lower than achieved in previous experiments. 
The energy spectrum between 180~GeV and 2~TeV 
is compatible with a power law of slope -2.72$\pm$0.14.
A crude variability analysis over the period of observation has 
shown no significant variation 
of the $\gamma$ flux.  The quiescent spectrum can be considered 
to match the spectrum measured by HEGRA at higher energy 
during past periods of equally low X-ray activity.  
We therefore tentatively conclude that
a steady VHE emission component has been identified in the 
spectrum of 1ES1959+650.  

EGRET has observed \citep{strong} a diffuse $\gamma$ spectrum 
(a part of the extragalactic background light, or EBL) in the energy 
range 100~MeV - 100~GeV. The question arises if emission from multiple
VHE $\gamma$ sources of the 1ES1959 type in a low state of activity could 
produce such a spectrum, when extrapolated to higher energies. 
One can estimate that about 400 such sources would be needed.
No complete HBL catalogue for the entire sky
exists, but this number of HBLs at redshift $z<0.5$ can be expected from 
the EMSS \citep{einst} on a purely statistical basis.  
However, their mean X-ray
flux is much lower than that of 1ES1959+650. Adopting the observed  
VHE/X-ray spectral index of $\alpha_{\rm X\gamma}\simeq 1$
for the quiescent spectrum,
the cumulative VHE flux of HBLs would thus 
be insufficient to explain the observed diffuse $\gamma$ spectrum.
In agreement with this finding, the sky positions of high energy photons 
($>10$~GeV) observed with EGRET
do not generally coincide with the positions of HBLs \citep{thompson},
indicating that other contributions to the EBL may also be important 
\citep{elsaesser}. In this context it should be noted that more measurements
of other $\gamma$-emitting AGNs during periods of low X-ray activity would
be important to shed light on these questions.

Explaining the observed quiescence spectrum 
by a one-zone SSC model is possible, but with some difficulty, 
since the implied relativistic electron pressure
exceeds the magnetic pressure, leading to an unstable situation. 
SSC models, on the other hand, clearly fall short of explaining the orphan 
flares seen in previous observations of 1ES1959+650
\citep{multiwl,sedwhipple,tonello1}. The quiescence spectrum and the
flares both seem to indicate the presence of an additional 
high energy electron population, possibly of hadronic origin 
\citep{boettcher,mannheim,massaro}, 
or proton synchrotron radiation \citep{aharonian}.
Short variability time scales might reflect
dynamical effects in shock-in-jet models,
or the short cooling times of protons at
ultra-high (up to 10$^{19}$~eV) energies \citep{rachen}.

Multi-wavelength monitoring campaigns are required to further reveal 
the nature of the VHE emission component in 1ES1959+650.
Such monitoring should also include future large neutrino 
observatories: the models based on  the presence of a 
significant hadronic component of the 
1ES1959+650 jet, e.g.~\citep{boettcher}, 
predict in a natural way also detectable neutrino fluxes.
The AMANDA collaboration, operating a neutrino telescope 
in the southern hemisphere,
recently reported five recorded neutrino events from 
the direction of 1ES1959+650,
over a total observation period of four years \citep{amanda}.
Three events coincided with 1ES1959+650 flares; 
one is coincident with the orphan flare
observed by the Whipple collaboration.   
While these observations are tantalizing, but
not yet statistically compelling, they do demonstrate 
that neutrino astronomy has reached
the stage at which fluxes at the level of the $\gamma$ 
fluxes observed with IACTs 
can be probed. Even for neutrino-to-$\gamma$ 
ratios smaller than unity,
ICECUBE should soon provide the necessary 
experimental sensitivity \citep{hh}.

\section*{Acknowledgements}

We would like to thank the IAC for the excellent working conditions on 
the La Palma Observatory Roque de los Muchachos. The
support of the German BMBF and MPG, the Italian INFN and the Spanish CICYT
is gratefully acknowledged. This work was also supported by ETH Research Grant 
TH-34/04-3.

\end{document}